# Enhancement of the UV emission from gold/ZnO nanorods exhibiting no green luminescence.

*Saskia Fiedler\*[‡¥], Laurent O. Lee Cheong Lem[‡ŧ], Cuong Ton-That[‡], Axel Hoffmann[†], Matthew R. Phillips\*[‡]*

[‡]School of Mathematical and Physical Sciences, University of Technology Sydney, 15 Broadway, Ultimo NSW 2007, Australia

[¥]Centre for Nano Optics, University of Southern Denmark, Campusvej 55, 5230 Odense M, Denmark

[ŧ]Australian National Fabrication Facility, Australian National University, Canberra ACT 2601, Australia

[†]Technische Universität Berlin, Fakultät II, Institut für Festkörperphysik, Sekretariat EW 5-4, Hardenbergstr. 36, 10623 Berlin, Germany

**Abstract**

Large reflection losses at interfaces in light emitting semiconductor devices cause a significant reduction in their light emission and energy efficiencies. Metal nanoparticle (NP) surface coatings have been demonstrated to increase the light extraction efficiency from planar high refractive index semiconductor surfaces. This emission enhancement in Au NP-coated ZnO is widely attributed to involvement of a green (~ 2.5 eV) deep level ZnO defect exciting localized surface plasmons in the NPs that non-radiatively decay into hot electrons, which return to ZnO and radiatively recombine. In this work, we achieve a 6 times enhancement of the ultra-violet excitonic emission in ZnO nanorods that are coated with 5 nm Au NPs without the aid of ZnO defects. Cathodoluminescence (CL), photoluminescence (PL) and a novel concurrent CL-PL technique as well as time resolved PL spectroscopy revealed that the increase in UV emission is due to the formation of an additional fast excitonic relaxation pathway that increases the exciton spontaneous emission rate. Concurrent CL-PL measurements ruled out the presence of charge transfer mechanism in the emission enhancement process. While time resolved PL results confirmed the existence of a new excitonic recombination channel that is attributed to exciton relaxation via the plasmon-assisted excitation of rapid non-radiative Au interband transitions. Our results establish that ZnO defect levels ~ 2.5 eV are not required to facilitate Au NP induced enhancement of the ZnO UV emission.

**Introduction**

Light emitting diodes (LEDs) fabricated from nanorods have clear and significant advantages over conventional planar device structures. These benefits include, a vast junction area due to the nanorods' size and three-dimensional morphology, the potential for light waveguiding, fabrication of films with low defect densities and strain as well as excellent carrier confinement. Zinc oxide (ZnO) nanorods are particularly attractive for LED applications owing to their: (i) attractive optical, electrical and mechanical properties, (ii) large surface to volume ratio, (iii) availability in a large assortment of bespoke shapes and sizes and (iv) facile growth on a wide variety of substrates.[1–3] Furthermore, because ZnO has a direct wide band gap at room temperature ($E_g = 3.37$ eV) as well as a large exciton binding energy of 60 meV ZnO, it is a very promising material for the development of ultra-violet (UV) LEDs.[4–8] However, despite the high luminescence efficiency of the near band edge (NBE) in ZnO, only a small fraction of this light generated is emitted due to large internal surface reflection losses arising from the high refractive index difference at the air – ZnO interface. Recently, however, it has been established that this optical limitation on the light extraction efficiency can be overcome by using a nanostructured gold thin-film surface coating, which has been found to significantly enhance that ZnO NBE emission output.[9–28]

Increase in the ZnO NBE luminescence intensity due to Au nanoparticle (NP) surface coatings have been attributed to the formation of an additional fast relaxation channel due to dipole-dipole coupling between excitons and NP plasmon modes, which increases the spontaneous emission rate (SER).[29–31] However, this mechanism is unlikely to be an efficient process for gold/ZnO systems because of the large energy difference between the ZnO exciton UV NBE emission at around 3.37 eV and the Au NP longitudinal surface plasmon (LSP) resonance ~ 2.5 eV. Accordingly, two different alternate models have been proposed both involving a deep level at ~ 2.5 eV below the ZnO conduction band (CB) and the charge transfer (CT) of hot electrons between ZnO and Au NPs as illustrated in Fig 1.[10,15,25,26] In the first model, a strong green luminescence (GL) generated from the ZnO by native surface defects is resonantly absorbed by the Au NPs. Following excitation, the LSPs then decay into hot carriers, in particular, hot electrons than transfer from high energy Au electronic levels into energetic ZnO CB states, which thermalize to the band edge. The subsequent radiative recombination of these CB electrons with free holes in the valence band (VB) enhances the UV excitonic emission at the expense of the GL; this mechanism has been observed experimentally for gold/ZnO. In the second CT model, electrons trapped at ZnO defects are transferred to the Au NPs due to their close energy alignment: ZnO surface defect electronic levels and the Fermi level of the Au NPs are located at ~- 5.35 eV and ~-5.30 eV below the vacuum level, respectively.[26,32] This CT process raises the electron density in Au NPs, forcing electrons into upper energy levels. These hot carriers in the NPs subsequently flow into the ZnO CB where they can relax radiatively and produce an increase in the UV emission.

The viability of both these CT models, as illustrated in Fig.1, are contingent on the existence of a deep-level ZnO emission band that resonates with the LSPs of metallic coatings. Here, we demonstrate that an increase of the UV NBE emission in Au NP-coated ZnO NR samples can be achieved without the presence of a resonant ZnO emission. Clearly neither of the currently proposed CT models outlined above can be used to describe the increased light output as they both rely on the presence of a strong ZnO green surface defect luminescence to produce the enhancement of the UV emission. Consequently, an alternate explanation is presented in this paper. Here, the enhanced UV emission in Au NP coated ZnO is attributed to the creation of an additional and faster non-radiative exciton relaxation pathway involving the excitation of interband Au transitions that increases the measured UV emission output by raising the UV excitonic SER. In this work, the UV emission enhancement in ZnO nanorods without a green defect luminescence has been systematically studied using cathodoluminescence (CL) and photoluminescence (PL) spectroscopy. We report a 6-fold enhanced UV emission from ZnO NRs, which exhibit negligible GL, by coating these NRs with a uniform surface coating of 5 nm Au NPs. A shortened UV radiative recombination exciton PL life time and data from a concurrent CL-PL technique confirm that an increase in the spontaneous emission rate (SER) is the underlying enhancement mechanism, following the creation of an additional exciton recombination pathway due to plasmon assisted fast Au NP interband transitions. Our results establish that the presence of a deep defect ZnO level is not required to facilitate the UV emission enhancement from ZnO nanorods with a 5 nm surface coating.

**Experimental Methods**

The ZnO NRs were prepared by a low temperature hydrothermal method.[33] A ZnO seed layer was deposited on a silicon wafer by drop casting zinc acetate (>98%, $(C_2H_3O_2)_2Zn*2H_2O$) solution (5 mM in ethanol) which was subsequently heated to 250°C in air for 20 minutes. The ZnO NRs were grown by placing the ZnO seed layer-coated silicon substrate in an autoclave containing a mixture of 25 mM zinc nitrate hexahydrate (98%, $Zn(NO_3)_2 6H_2O$), and 25 mM hexamethylenetetramine (HMT, ≥99.0%, $C_6H_{12}N_4$) from Sigma-Aldrich. The autoclave was held at 90°C for 3 hours before removing the substrates and rinsing them in de-ionized water. Uniform Au nanoparticles (NPs) were formed by sputtering an Au thin film with a nominal thickness of one nanometer followed by annealing at 300°C for 30 minutes in air. The same deposition process was used to coat a polished both sides *a*-plane 5 x 5 x 1 mm ZnO single crystals plate (MTI Corp.) which was used as a reference sample.

The morphology and size of the ZnO NR samples before and after the deposition of the Au NP coating were studied with a Field Emission Scanning Electron Microscope (Zeiss Supra 55VP). Optical transmission spectra were collected with an integrated sphere connected to an Ocean Optics QE Pro spectrometer. CL spectroscopy was performed in an FEI Quanta 200 SEM equipped with a liquid helium and liquid nitrogen cold stage. Light emitted from the sample was collected by a parabolic mirror and analyzed using an Ocean Optics QE Pro spectrometer. Light injection in the CL chamber enabled concurrent CL and PL spectroscopy of spatially equivalent regions of the samples, where the parabolic mirror was used to focus the laser light onto the surface of the sample in addition to collecting the photoluminescence and cathodoluminescence. Sub-bandgap excitation at $\lambda = 532$ nm (Lambda Pro laser 20 mW) allowed to study the possible CT mechanism in the Au NP-ZnO NR samples. All luminescence data were corrected for the total system response. The dynamic behavior of the charge carriers in the Au-coated ZnO NRs was investigated by time-resolved PL (TR-PL) using a pulsed fiber laser for excitation, with an excitation wavelength of $\lambda = 258$ nm, pulse repetition rate of 76 MHz and pulse duration of approximately 5.5 ps.

**Results and discussion**

**Structural Properties of ZnO nanorods coated with Au nanoparticles**

The low temperature hydrothermal growth produced a continuous and dense ZnO NR surface film. The hexagonal ZnO NRs with a growth axis along the <0001> direction exhibit an average diameter of $40 \pm 10$ nm and length of around 700 nm, which are oriented at different angles to the normal direction of the Si substrate (fig. 2a). Fig. 2b shows the ZnO NR's surface coating consisting of uniformly distributed Au NPs with an average diameter of 5 nm. Only half of the ZnO NR sample was coated with Au NPs leaving the other, uncoated half as a reference to compare the CL and PL measurements before and after the deposition of the metal NP coating.

**Optical Properties of Au NP-coated ZnO NRs**

Typical optical transmission spectra from the ZnO single crystal plate reference sample before and after the deposition of a 5 nm Au NP surface coating are shown in Fig. 3. The spectrum from the uncoated ZnO sample reveal a strong decrease in the UV transmission due to the band edge absorption, as expected. After the deposition of the 5 nm Au NP coating, the transmission spectrum displayed an additional broad absorption band centered at ~ 2.25 eV with an extended absorption tail in the UV spectral region. The dip in the green spectral region is characteristic of the LSP plasmon resonance absorption of spherical Au NPs with a diameter of 5 nm and confirms the excitation of LSP modes in the Au NPs.[34–36]

A comparison of typical normalized CL (5 kV) and PL (excitation at $\lambda_{exc}$ = 325 nm) spectra at 80 K of an uncoated ZnO nanorod ensemble is shown in Fig. 4. The luminescence spectra consist of two emission peaks centered at 3.33 and 1.75 eV. Significantly, no GL was observed in any of the ZnO nanorod surface coatings. The NBE UV emission is due to radiative recombination of free excitons (FX) and their phonon replica which produce a low energy tail. While the relatively weaker, broad deep level (DL) red emission centered at ~ 1.7 eV has been assigned to native point defects,[37–40] the RL intensity is noticeably stronger in the PL spectrum compared with its CL counterpart. This result is in agreement with a near surface distribution of radiative DL centers in the ZnO NRs: The PL excitation is strongest at the NR surface as the optical absorption follows Beer's law, whereas the maximum CL excitation at 5 kV occurs deeper in the core of the NRs.

To investigate the effects of the 5 nm Au NP coating on the light emission output from ZnO NRs, CL spectroscopy was carried out at different temperatures with an accelerating voltage of 5 kV, providing a similar probing depth to the PL measurements. Fig. 5 shows the CL spectra of annealed uncoated

ZnO NRs and 5 nm Au NP coated ZnO NRs at a temperature of 10 K and 80 K. The CL spectra are dominated by the ZnO NBE UV emission peak at ~ 3.35 eV which is attributed to (i) recombination of bound excitons (BX) at 10 K and (ii) free excitons (FX) at 80 K, where the bound excitons have thermally dissociated. Comparison of the CL spectra of the ZnO NRs with and without Au NPs, displayed in Fig. 5, reveals that the Au NP coating produces a 6 times and 1.2 times increase in the integrated NBE emission intensity at 80 K and 10 K respectively, while the DL emission intensity remains unchanged. A higher UV emission enhancement when raising the temperature from 10K to 80K is consistent with an excitonic coupling mechanism. FX dominate the emission at 80 K and are expected to interact efficiently with the Au NP surface coating due to their high mobility and greater spatial extent. Conversely, at 10K, BX prevail and as they are spatially localized the exciton coupling strength is reduced.

Concurrent excitation of the Au NP coated ZnO NRs with a green laser ($\lambda_{exc}$ = 532 nm) and electron beam was carried out to investigate the CT mechanism where hot electrons in Au NPs could be responsible for the increase in the UV ZnO emission, as described above. If this mechanism is correct, it is expected that under the green laser illumination alone, only LSPs in the Au NPs would be excited due to laser energy of 2.33 eV being very close to the Au NP plasmon resonance ($\hbar\omega_{LSP}$ ~ 2.25 eV) while no excitons are created in the ZnO as the light excitation is sub-band gap. However, simultaneous excitation with the electron beam and the laser should induce both, UV and DL emission of the ZnO NRs, and the LSPs in the Au NPs.

Fig.6 (a) shows the luminescence spectra of the Au NP coated ZnO sample at 10 K using excitation from the electron beam only, the green laser only, and both the green laser and the electron beam concurrently. Electron beam only excites an intense UV emission (~ 3.34 eV) and a weak broad red DL emission centered at 1.7 eV, attributed to bound excitons and native defects, respectively. Illumination with the green laser only produces a broad luminescence band centered 2.0 eV, and significantly no NBE UV emission is observed. The orange luminescence (OL) has been assigned to sub-band gap excitation of ionized acceptors in ZnO. In this process, the laser excites an electron at an ionized acceptor into the conduction band forming a neutral acceptor. The subsequent radiative recombination of the excited conduction band electron with the neutral acceptor reforms the initial acceptor state generating the OL.[41] It is also important to note that the defect level of the orange emission is likely to be at a higher energy in the band gap closer to the Au NP Fermi level because it deep level centers in ZnO exhibit a large Stokes shift between their excitation and emission energies.[42] In addition, visible PL from Au NPs has been reported, however, the 5 nm Au NPs used in this work are too large to enable the luminescence recombination mechanism, as discussed below.

To establish whether the presence of free holes is required to facilitate the UV emission enhancement via the CT process as described above, luminescence spectra were collected from the Au NP coated ZnO under coincident laser and electron beam excitation. Figure 6 (b) shows that there is no perceivable change in the intensity or shape of the NBE emission at 5 kV when comparing luminescence spectra using electron beam only illumination to those measured with concurrent green laser and electron beam excitation. These results indicate that the Au NP surface coating induced ZnO UV NBE enhancement observed in this work does not originate from a CT mechanism as reported in other studies

A typical high-resolution PL spectrum of the UV NBE of the ZnO NRs is displayed in Fig. 6(a). The PL spectrum is dominated by donor bound $I$ line transitions (DBX) around 3.36 eV with their LO-phonon replica ($\hbar\omega_{LO} = 73$ meV) and a peak at 3.32 eV is due to the overlap of the two-electron satellite (TES) of the DBX and the 1-LO-FX emission peaks. A weak emission peak around 3.26 eV is attributed to the 2-LO FX transition. These DBX peaks are much broader than those reported in ZnO single crystals most likely due to the poorer crystal quality of the hydrothermally-produced ZnO NRs.[43] Consequently, the 3.36 eV peak is too broad to allow its assignment to a specific DBX $I$ line transition.

The enhancement factor as a function of energy, as shown in Fig 6(b) is obtained by dividing the PL spectra of the ZnO NRs with and without the Au NP coating. A typical ratio plot, shown in Fig. 6(b), reveals that the energies at which the maximum enhancement occurs are highly correlated with BX and FX emission peak positions. This result confirms that the Au NP surface coating specifically enhances the excitonic emission peaks from the ZnO NRs rather than uniformly increasing the NBE emission.

To establish the effect of the 5 nm Au NP surface coating on ZnO NRs on the radiative lifetime of the ZnO NBE emission, TR-PL was collected at a fixed wavelength of λ = 368 nm at a temperature of 8 K (fig. 7). The TR-PL curve for both the uncoated and Au NP coated ZnO NR samples was de-convoluted with the system response and fitted with a bi-exponential function. The resulting two time constants were averaged over five positions on each sample to minimize possible local variations in homogeneity.

The annealed uncoated ZnO NRs exhibited a short life time component of $\tau_{NR} = (25.2 \pm 3.2)$ ps, which was attributed to non-radiative recombination.[29] While the longer life time component of $\tau_R = (129.4 \pm 4.5)$ ps is due to the radiative excitonic recombination, where the total relaxation time τ is given by : $\frac{1}{\tau} = \frac{1}{\tau_R} + \frac{1}{\tau_{NR}}$.[29,44,45] The ZnO NRs decorated with Au NPs showed a similar non-radiative life time of $\tau_{NR}^* = (21.5 \pm 1.5)$ ps, suggesting that the fast component is due to non-radiative recombination in the bulk rather than at surface states, which are likely to be passivated by the metal surface coating. This conclusion is consistent with the CL results which revealed that there was no change in the ZnO DL emission due to the Au NP surface coating (cf. Fig. 4). Conversely, the radiative life time of Au NP coated ZnO NRs was reduced to $\tau_R^* = (91.0 \pm 15.2)$ ps. A shorter UV NBE life time, τ*$_{Au}$, due to the Au NP surface coating is generally considered to be strong evidence for the formation of an additional, faster exciton recombination channel, which enhances the SER of

the ZnO NRs with

$$\frac{1}{\tau^*} = \frac{1}{\tau_R^*} + \frac{1}{\tau_{NR}^*} + \frac{1}{\tau_{Au}^*}.$$ [29,30,46]

As described above, in ZnO samples with ~ 2.5 eV deep level defects, the additional, faster exciton relaxation pathway has been attributed to the excitation of LSPs in the Au NPs facilitated by the presence of a GL that strongly overlaps with the Au NP LSP absorption energy.[9,17,24,47–49] However, since this green ~ 2.5 eV defect center is absent in the samples studied in this work, a different model is required to explain the observation of the enhanced UV NBE in the Au NP coated ZnO NRs. As demonstrated above, a CT mechanism can be also excluded as there is no hot electron transfer from the Au nanoparticles into the conduction band of the ZnO NRs under simultaneous electron beam and green laser ($\lambda_{exc}$ = 532 nm). Furthermore, since there is no strong spectral overlap between the LSP absorption resonance ($E_{LSP}$ = 2.25 eV) in the Au NPs with either red defect center ($E_{Rl}$ = 1.75 eV) or UV NBE ($E_{exciton}$ = 3.36 eV) emissions from the ZnO NRs, exciton relaxation cannot occur via a radiationless exciton-LSP coupling mechanism. Nevertheless, as evidenced by a reduced radiative lifetime of the exciton emission, an additional, faster exciton decay channel is indeed created by the 5 nm Au NP surface coating.

The two key criteria that must be met for this new relaxation pathway to produce an enhancement of the ZnO excitonic SER are: (1) its energy is close to that of the excitonic emission of ZnO (3.36 eV) and (2) the relaxation channel is faster than the excitonic recombination in ZnO. The interband transitions in Au can fulfil both of these requirements. Electrons in Au can be promoted via interband transitions from occupied 5d levels to unoccupied states in hybridized 6sp bands above the Fermi level with an excitation threshold energy of ~ 2.4 eV.[50,51] These generated hot electrons and holes rapidly thermalize and non-radiatively recombine via electron-electron and electron-phonon scattering mechanisms with a fast sub-ps relaxation time.[52] Indeed, it is noteworthy that an increase of the UV photo-response in Au-ZnO nanocomposite sensors is attributed to Au interband transitions increasing the photo-conductivity of the device, which is in agreement with our proposed mechanism for the additional relaxation channel.[22]

In Au NPs, due to the carrier confinement effects reducing the density of states (DOS), this non-radiative Au channel slows down, facilitating a radiative relaxation involving an interband transition around the L- and X- symmetry point in the Au band structure. However, as the 5 nm Au NPs used in this work are sufficiently large enough to exhibit a bulk-like DOS, these luminescence interband Au transitions cannot occur.[53] Furthermore, no additional signature Au NP luminescence emission was observed in any of the PL measurements, confirming that the non-radiative sub-ps interband transitions within the 5 nm Au NPs are many orders of magnitude faster that the radiative decay time for DBX (~ 100s ps) and FX (~ 1 - 10s ns) in ZnO.[43,54] Accordingly, the Au NPs facilitate an additional fast excitonic relaxation pathway involving the excitation of Au interband transitions via non-radiative energy transfer process, which increases the exciton SER and enhances the UV NBE emission output. This is a significant finding as it reveals the importance of the Au NP size on the observed emission enhancement. The Au NP needs to small enough to support LSP generation but sufficiently large to exhibit bulk-like properties. Furthermore, the NPs cannot be too small (< 1 nm) since the quantum confinement effect will reduce the DOS of the AU NPs and slow down relaxation rate, as discussed above.

It is also important to note that luminescence spectra (data not shown) from an as-deposited 1 nm thick featureless Au surface film before heating to form the 5nm NPs, exhibit a small decrease in NBE output intensity when compared to its uncoated reference sample rather than an increase. This result suggests that a nanostructured Au coating must be capable exciting LSPs to produce the observed enhancement of the UV ZnO NBE emission intensity. It is likely that the additional highly efficient, fast recombination channel required to increase the exciton SER involving holes in the 5d bands with electrons in 6sp states is assisted by excitation of surface plasmons.[55]

**Conclusion**

Au NP coated ZnO nanorods with no green defect luminescence have been systematically studied using CL, PL and a novel concurrent CL-PL technique as well as time resolved PL spectroscopy. It was shown that the UV NBE emission of the ZnO NRs is increased up to 6 times by the Au NP surface coating despite the absence of a ZnO defect level to excite LSPs in the Au NPs. Concurrent illumination with electron beam and green laser excited the LSPs in the Au NPs but no increase of the UV emission was observed, ruling out a CT mechanism as the cause of the observed NBE enhancement. A shortened NBE PL life time from ZnO NRs with the 5 nm Au NP surface coating indicated the formation of an additional, faster relaxation channel that increased the exciton SER. This new exciton recombination pathway was attributed to relaxation via the plasmon-assisted excitation of rapid Au interband transitions. A ZnO nanorod deep level defect responsible for a RL was unaffected by the Au surface coating and played no role in the enhancement of the NBE emission. The finding that the size of the Au NP plays a crucial role on the emission enhancement factor is important for the utilization of Au NP surface coatings to improve the performance and energy efficiency of solid state LED lighting devices.

FIGURES.

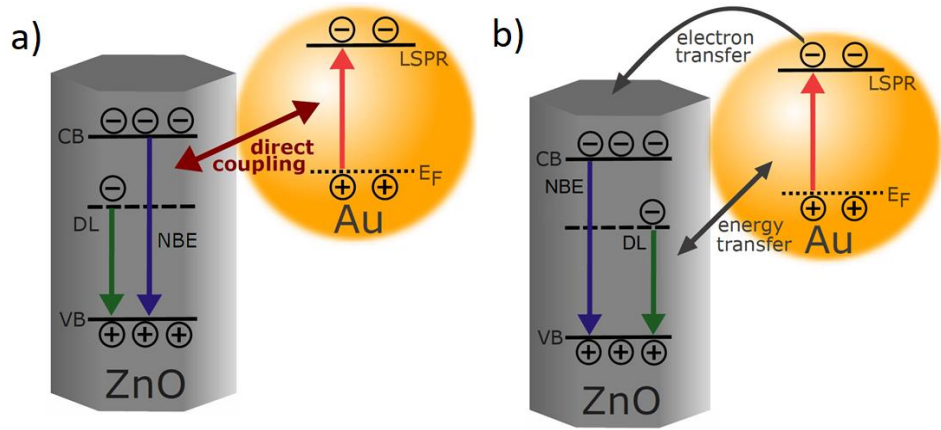

**Figure 1**: (a) Exciton-LSP coupling: direct dipole-dipole coupling between exciton in ZnO and LSPs in Au NPs which increases the SER enhancing the excitonic UV emission intensity from a ZnO nanowire. This mechanism is unlikely in Au NP coated ZnO nanorods due to the large difference between the exciton and Au NP LSP resonance energies. (b) Two previously proposed CT mechanisms: (i) green luminescence from ZnO defect levels is absorbed by Au NPs or (ii) electrons in defect level can transfer to Au. The LSPs produced in either process decay into hot carriers where hot Au electrons can flow into conduction band of ZnO and recombine with holes in ZnO valence band enhancing the UV emission intensity.

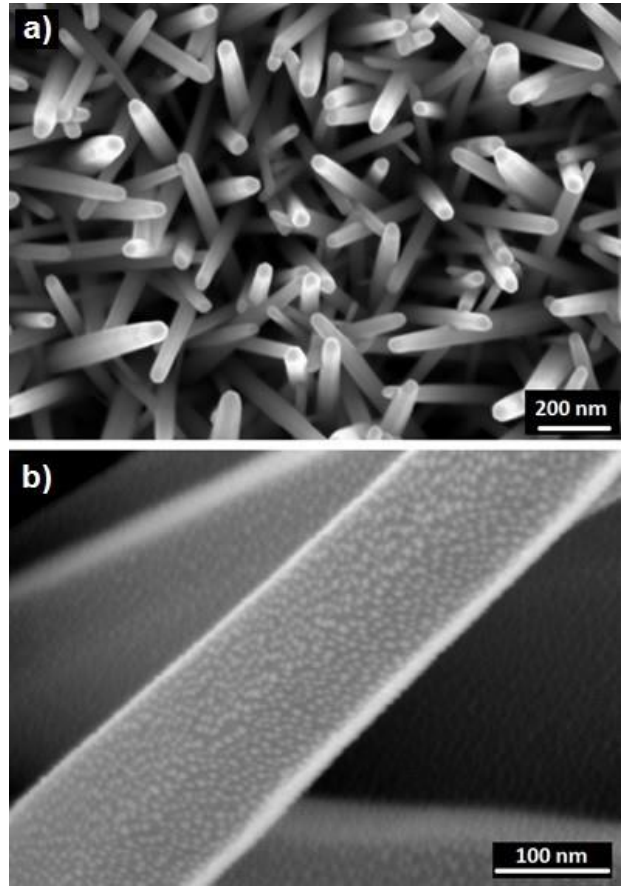

**Figure 2:** SE image of (a) as-grown hexagonal ZnO NRs grown on a Si substrate with an approximate diameter of 40 ± 10 nm, (b) ZnO NR decorated with uniform Au NP film with a relatively uniform diameter of 5 nm.

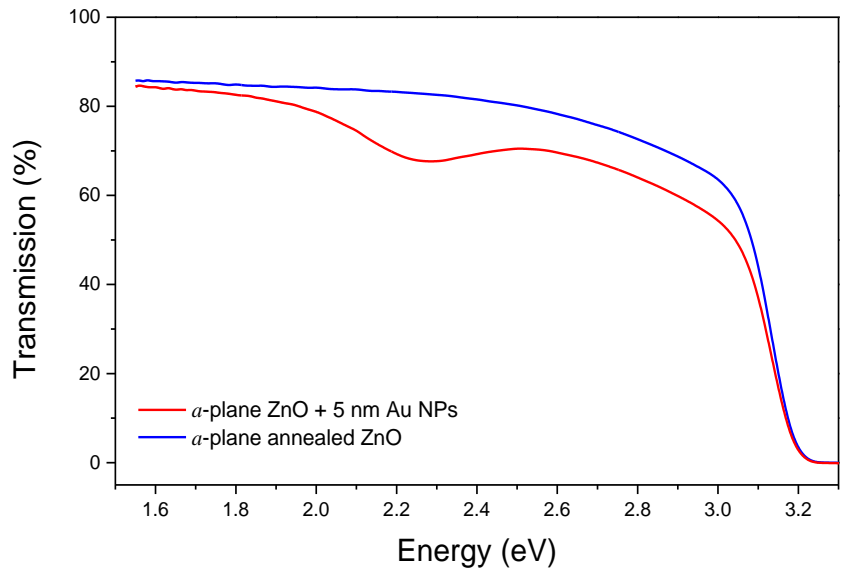

**Figure 3**: Transmission optical spectra of an annealed uncoated *a*-plane annealed ZnO single crystal plate (blue dashed) and *a*-plane ZnO crystal decorated with 5 nm Au NPs (red), showing the typical plasmon resonance absorption around 2.25 eV characteristics of the LSP resonance of 5 nm spherical Au NPs.

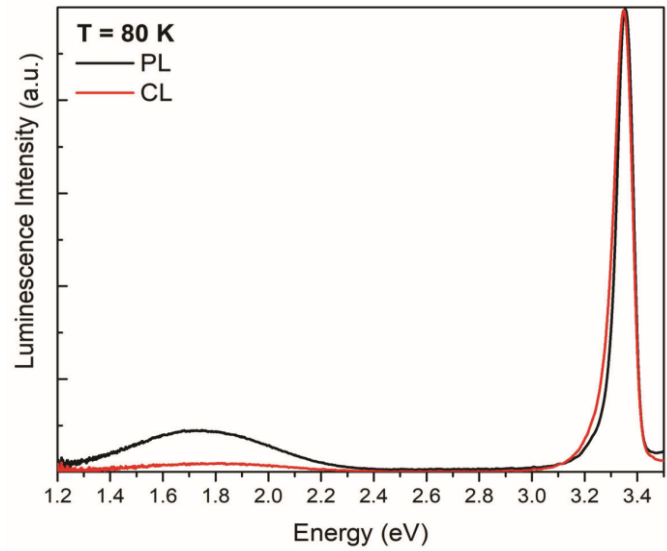

**Figure 4**: Normalized CL and PL spectra of the uncoated ZnO NRs at 80 K, confirming that the RL is most intense at the ZnO surface. CL: HV = 5 kV, P = 45 µW, scan area 15 µm × 15 µm. PL: $\lambda_{exc}$ = 325 nm, P = 35 µW, spot size ∼ 30 µm.

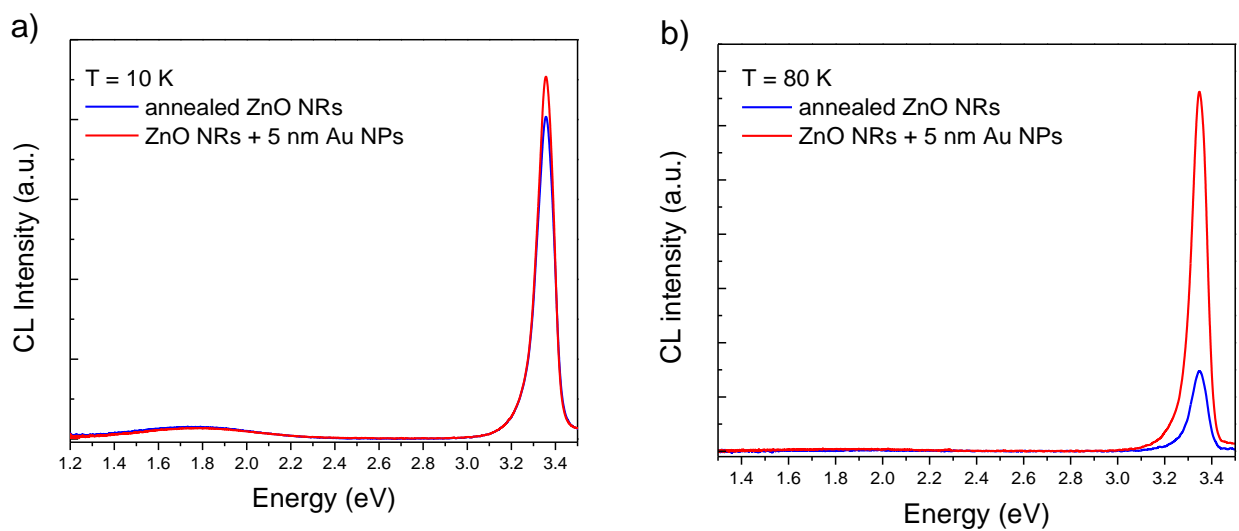

**Figure 5**: CL (HV = 5 kV, $I_b$ = 3.5 nA, scan 10μm x 10μm) spectra of annealed uncoated ZnO NRs (blue) and ZnO NRs with a surface coating of 5 nm Au NPs (red) showing (a) an enhanced UV emission at T = 10 K and (b) a 6-fold enhancement at T = 80 K due to Au NP coating. The intensity and shape of the RL is the same with and without the Au NP coating.

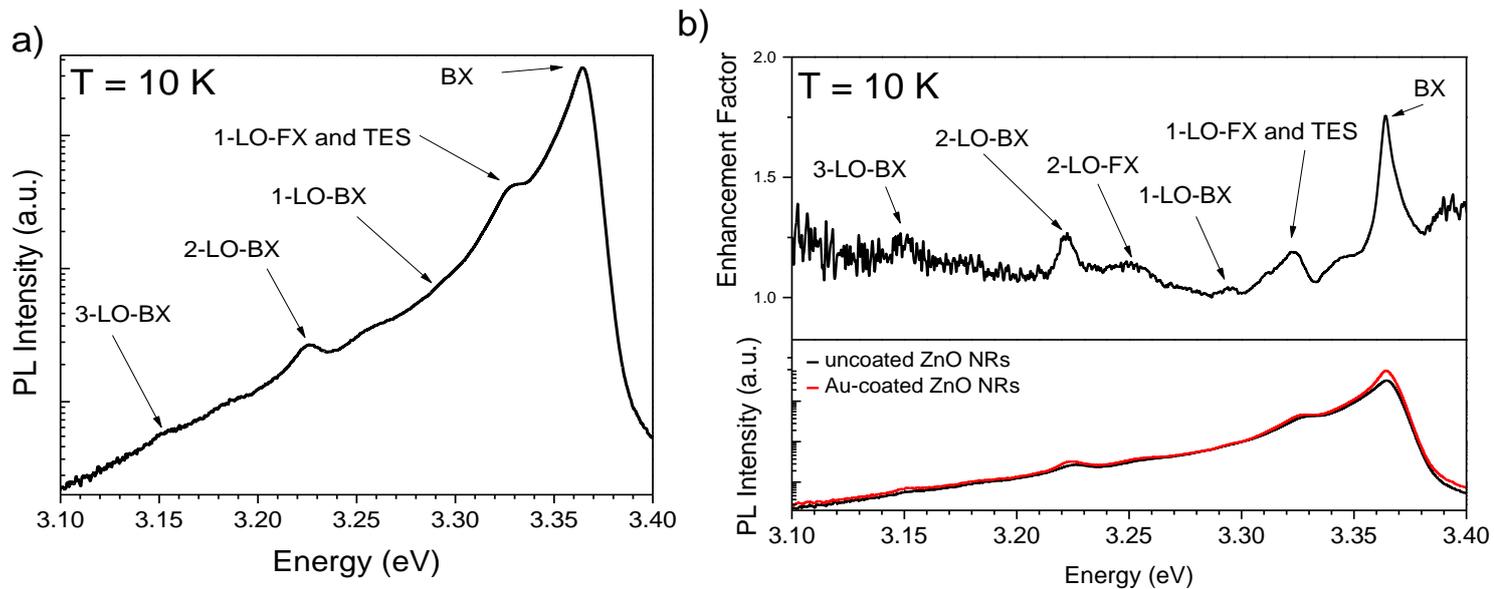

**Figure 6:** (a): High-resolution PL spectrum of NBE of the ZnO NRs at T =10 K dominated by DBX peak and its phonon replicas and TES transition. (b) Top: 10 K-PL enhancement factor of Au NP coated ZnO nanorods as a function of energy. Bottom: high-resolution PL of uncoated (black) and Au nanoparticle coated ZnO nanorods (red), graphed on a semi-logarithmic scale. The ratio of PL spectra with and without the Au NP surface coating provide the enhancement factor data. Excitation: $\lambda_{exc}$ = 325 nm and P = 22.4 mW, spot size ~ 30 μm.

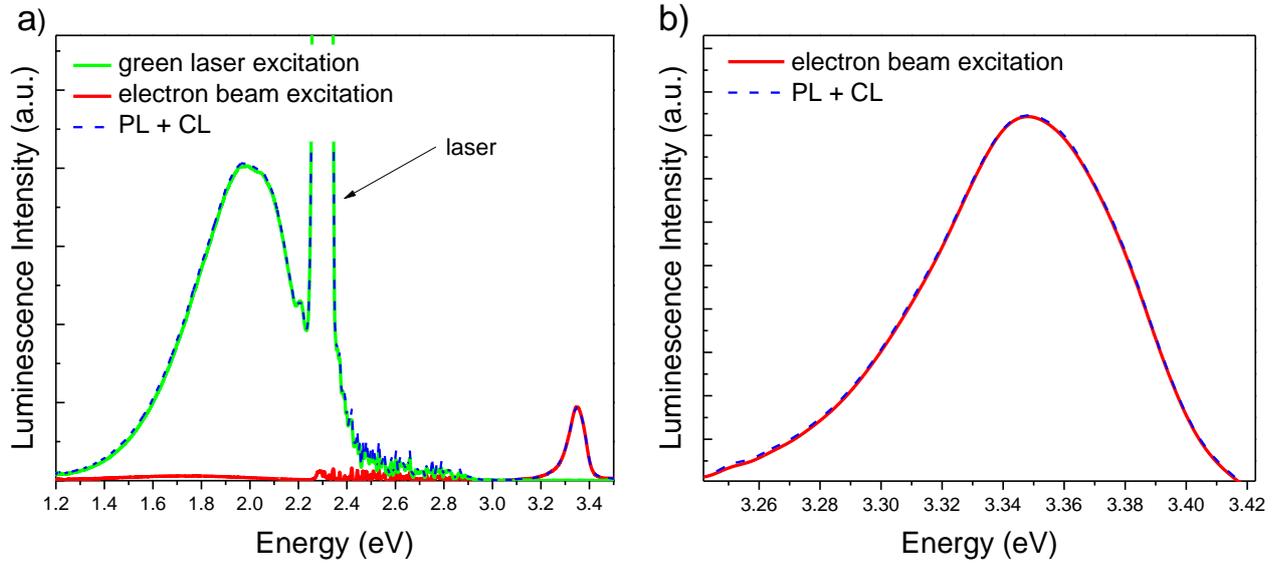

**Figure 6:** Luminescence spectra of ZnO NRs decorated with 5 nm Au NPs at T = 10 K. (a) A typical PL spectrum using green laser ($\lambda_{exc}$ = 532 nm) sub-band gap illumination (green full line) showing a broad OL peak centered at 2.0 eV emissions attributed to excitation and relaxation of ionized acceptors in ZnO. The intense PL emission at 2.3 eV is due to the green laser illumination. A CL spectrum at HV = 5 kV and $I_b$ = 3.5 nA (red full line) reveals a weak DL emission at 1.75 eV in the visible and a strong NBE emission at 3.34 eV attributed to BX. A luminescence spectrum (blue dashed line) using concurrent PL and CL excitation exhibiting a spectrum identical to the sum of the PL only illumination and CL excitation only spectra. (b) UV NBE emission spectrum using electron beam excitation only (red full line) and concurrent electron beam and green laser excitation (blue dashed line) showing identical emission spectra, indicating that a CT mechanism is not responsible for the enhanced UV NBE in ZnO coated with Au NPs.

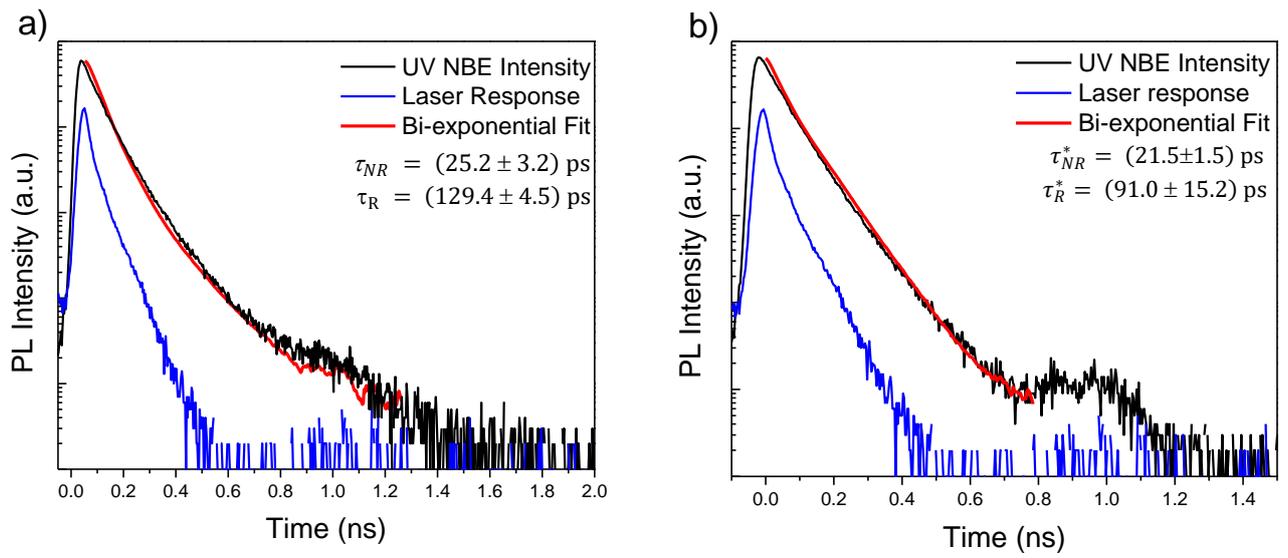

**Figure 7**: Typical time-resolved UV NBE PL (T = 8 K) of (a) annealed uncoated ZnO NRs and (b) ZnO NRs decorated with a surface coating of 5 nm Au NPs, showing a shorter UV NBE radiative recombination life time for the Au NP coated ZnO NR sample. This result provides evidence for the creation of an additional, fast ZnO exciton decay channel due to the Au NPs coating.

AUTHOR INFORMATION

**Corresponding Authors**

*safi@mci.sdu.dk

*matthew.phillips@uts.edu.au

**Author Contributions**

The manuscript was written through contributions of all authors. All authors have given approval to the final version of the manuscript.

**Competing Interests**

The authors declare no competing interests.

ACKNOWLEDGEMENT.

This work was supported by the Australian Research Council (DP150103317). The authors gratefully acknowledge the technical assistance of the staff at the Microstructural Analysis Unit at University of Technology Sydney, in particular Geoffrey McCredie and Katie McBean.

REFERENCES.

1   G. Amin, PhD thesis, Linköpings Universitet, 2012.
2   N. Bano, S. Zaman, A. Zainelabdin, S. Hussain, I. Hussain, O. Nur and M. Willander, *J. Appl. Phys.*, 2010, **108**, 043103.
3   A. Manekkathodi, M.-Y. Lu, C. W. Wang and L.-J. Chen, *Adv. Mater.*, 2010, **22**, 4059–4063.
4   H. Li, Y. Huang, Q. Zhang, J. Liu and Y. Zhang, *Solid State Sci.*, 2011, **13**, 658–661.
5   P. Miller, PhD thesis, Univeristy of Canterbury, 2008.
6   Ü. Özgür, Ya. I. Alivov, C. Liu, A. Teke, M. A. Reshchikov, S. Doğan, V. Avrutin, S.-J. Cho and H. Morkoç, *J. Appl. Phys.*, 2005, **98**, 041301.
7   C. Soci, A. Zhang, B. Xiang, S. A. Dayeh, D. P. R. Aplin, J. Park, X. Y. Bao, Y. H. Lo and D. Wang, *Nano Lett.*, 2007, **7**, 1003–1009.
8   D. G. Thomas, *J. Phys. Chem. Solids*, 1960, **15**, 86–96.
9   Z. Yi, J. Chen, J. Luo, Y. Yi, X. Kang, X. Ye, P. Bi, X. Gao, Y. Yi and Y. Tang, *Plasmonics*, 2015, **10**, 1373-1380.
10  V. Perumal, U. Hashim, S. C. B. Gopinath, H. Rajintra Prasad, L. Wei-Wen, S. R. Balakrishnan, T. Vijayakumar and R. A. Rahim, *Nanoscale Res. Lett.*, 2016, **11**, 11:31.
11  G. Bertoni, F. Fabbri, M. Villani, L. Lazzarini, S. Turner, G. Van Tendeloo, D. Calestani, S. Gradečak, A. Zappettini and G. Salviati, *Sci. Rep.*, 2016, **6**, 19168.
12  W. Chamorro, J. Ghanbaja, Y. Battie, A. E. Naciri, F. Soldera, F. Muecklich and D. Horwat, *J Phys Chem C*, 2016, **120**, 29405–29413.
13  M. Mahanti and D. Basak, *Chem. Phys. Lett.*, 2014, **612**, 101–105.
14  L. Wang, X. Wang, S. Mao, H. Wu, X. Guo, Y. Ji and X. Han, *Nanoscale*, 2016, **8**, 4030.
15  C. W. Cheng, E. J. Sie, B. Liu, C. H. A. Huan, T. C. Sum, H. D. Sun and H. J. Fan, *Appl. Phys. Lett.*, 2010, **96**, 071107.
16  D.-R. Hang, S. E. Islam, C.-H. Chen and K. H. Sharma, *Chem. - Eur. J.*, 2016, **22**, 14950–14961.
17  H. Y. Lin, C. L. Cheng, Y. Y. Chou, L. L. Huang, Y. F. Chen and K. T. Tsen, *Opt. Express*, 2006, **14**, 2372–2379.
18  M. Liu, R. Chen, G. Adamo, K. F. MacDonald, E. J. Sie, T. C. Sum, N. I. Zheludev, H. Sun and H. J. Fan, *Nanophotonics*, 2013, **2**, 153-160.
19  E. J. Guidelli, O. Baffa and D. R. Clarke, *Sci. Rep.*, 2015, **5**, 14004.
20  D. Zhang, H. Ushita, P. Wang, C. Park, R. Murakami, S. Yang and X. Song, *Appl. Phys. Lett.*, 2013, **103**, 093114.
21  A. Pescaglini, A. Martín, D. Cammi, G. Juska, C. Ronning, E. Pelucchi and D. Iacopino, *Nano Lett.*, 2014, **14**, 6202–6209.
22  N. Gogurla, A. K. Sinha, S. Santra, S. Manna and S. K. Ray, *Sci. Rep.*, 2014, **4**, 6483.
23  Y. Zhang, X. Li and X. Ren, *Opt. Express*, 2009, **17**, 8735–8740.
24  S. T. Kochuveedu, J. H. Oh, Y. R. Do and D. H. Kim, *Chem. - Eur. J.*, 2012, **18**, 7467–7472.
25  T. Singh, D. K. Pandya and R. Singh, *Thin Solid Films*, 2012, **520**, 4646–4649.
26  L. Su and N. Qin, *Ceram. Int.*, 2015, **41**, 2673–2679.
27  S. G. Zhang, L. Wen, J. L. Li, F. L. Gao, X. W. Zhang, L. H. Li and G. Q. Li, *J. Phys. Appl. Phys.*, 2014, **47**, 495103.
28  V. Perumal, U. Hashim, S. C. Gopinath, R. Haarindraprasad, W.-W. Liu, P. Poopalan, S. R. Balakrishnan, V. Thivina and A. R. Ruslinda, *PloS One*, 2015, **10**, e0144964.
29  Q. J. Ren, S. Filippov, S. L. Chen, M. Devika, N. Koteeswara Reddy, C. W. Tu, W. M. Chen and I. A. Buyanova, *Nanotechnology*, 2012, **23**, 425201.

30 Y. Zang, X. He, J. Li, J. Yin, K. Li, C. Yue, Z. Wu, S. Wu and J. Kang, *Nanoscale*, 2013, **5**, 574–580.
31 J. Lu, J. Li, C. Xu, Y. Li, J. Dai, Y. Wang, Y. Lin and S. Wang, *ACS Appl. Mater. Interfaces*, 2014, **6**, 18301–18305.
32 B. G. Yacobi, *Semiconductor Materials: An Introduction to Basic Principles*, Springer Science & Business Media, 2003.
33 S. Anantachaisilp, S. M. Smith, C. Ton-That, T. Osotchan, A. R. Moon and M. R. Phillips, *J. Phys. Chem. C*, 2014, **118**, 27150–27156.
34 G. Kalyuzhny, A. Vaskevich, M. A. Schneeweiss and I. Rubinstein, *Chem.- Eur. J.*, 2002, **8**, 3849–3857.
35 J. Siegel, O. Lyutakov, V. Rybka, Z. Kolská and V. Švorčík, *Nanoscale Res. Lett.*, 2011, **6**, 9.
36 X. D. Li, T. P. Chen, Y. Liu and K. C. Leong, *Opt. Express*, 2014, **22**, 5124.
37 A. B. Djurišić, Y. H. Leung, K. H. Tam, Y. F. Hsu, L. Ding, W. K. Ge, Y. C. Zhong, K. S. Wong, W. K. Chan, H. L. Tam, K. W. Cheah, W. M. Kwok and D. L. Phillips, *Nanotechnology*, 2007, **18**, 095702.
38 A. Janotti and C. G. Van de Walle, *Phys. Rev. B*, 2007, **76**, 165202.
39 H. Berrezoug, A. E. Merad, M. Aillerie and A. Zerga, *Mater. Res. Express*, 2017, **4**, 035901.
40 S. Anantachaisilp, S. M. Smith, C. Ton-That, S. Pornsuwan, A. R. Moon, C. Nenstiel, A. Hoffmann and M. R. Phillips, *J. Lumin.*, 2015, **168**, 20–25.
41 S. Choi, M. R. Phillips, I. Aharonovich, S. Pornsuwan, B. C. C. Cowie and C. Ton-That, *Adv. Opt. Mater.*, 2015, **3**, 821–827.
42 M. D. McCluskey, C. D. Corolewski, J. Lv, M. C. Tarun, S. T. Teklemichael, E. D. Walter, M. G. Norton, K. W. Harrison and S. Ha, *J. Appl. Phys.*, 2015, **117**, 112802.
43 M. R. Wagner, G. Callsen, J. S. Reparaz, J.-H. Schulze, R. Kirste, M. Cobet, I. A. Ostapenko, S. Rodt, C. Nenstiel, M. Kaiser, A. Hoffmann, A. V. Rodina, M. R. Phillips, S. Lautenschläger, S. Eisermann and B. K. Meyer, *Phys. Rev. B*, 2011, **84**, 035313.
44 E. J. Guidelli, O. Baffa and D. R. Clarke, *Sci. Rep.*, 2015, **5**, 14004.
45 W. Liu, H. Xu, S. Yan, C. Zhang, L. Wang, C. Wang, L. Yang, X. Wang, L. Zhang, J. Wang and Y. Liu, *ACS Appl. Mater. Interfaces*, 2016, **8**, 1653–1660.
46 J. Lu, J. Li, C. Xu, Y. Li, J. Dai, Y. Wang, Y. Lin and S. Wang, *ACS Appl. Mater. Interfaces*, 2014, **6**, 18301–18305.
47 Jun-Dar Hwang, M. J. Lai, H. Z. Chen and M. C. Kao, *IEEE Photonics Technol. Lett.*, 2014, **26**, 1023–1026.
48 X. Huang, R. Chen, C. Zhang, J. Chai, S. Wang, D. Chi and S. J. Chua, *Adv. Opt. Mater.*, 2016, **4**, 960–966.
49 C. Zhang, C. E. Marvinney, H. Y. Xu, W. Z. Liu, C. L. Wang, L. X. Zhang, J. N. Wang, J. G. Ma and Y. C. Liu, *Nanoscale*, 2015, **7**, 1073–1080.
50 G. V. Hartland, *Chem. Rev.*, 2011, **111**, 3858–3887.
51 L. Le Thi Ngoc, J. Wiedemair, A. van den Berg and E. T. Carlen, *Opt. Express*, 2015, **23**, 5547–5564.
52 S. Link and M. A. El-Sayed, *Rev. Phys. Chem.*, 2000, **19**, 409–453.
53 E. Dulkeith, T. Niedereichholz, T. Klar, J. Feldmann, G. von Plessen, D. Gittins, K. Mayya and F. Caruso, *Phys. Rev. B*, 2004, **70**, 205424.
54 J. S. Reparaz, F. Güell, M. R. Wagner, A. Hoffmann, A. Cornet and J. R. Morante, *Appl Phys Lett*, 201, **4**, 053105.
55 T. V. Shahbazyan, I. E. Perakis and J.-Y. Bigot, *Phys. Rev. Lett.*, 1998, **81**, 3120–3123.